\documentclass[superscriptaddress,prd,aps,preprint]{revtex4-1}
\newcommand{\bea}{\begin{eqnarray}}
\newcommand{\eea}{\end{eqnarray}}

\usepackage{graphicx,amsfonts,amsmath}

\begin{document}

\title{On the superfield supersymmetric aether-like Lorentz-breaking models}

\author{C. F. Farias}
\affiliation{Departamento de F\'{\i}sica, Universidade Federal da Para\'{\i}ba\\
 Caixa Postal 5008, 58051-970, Jo\~ao Pessoa, Para\'{\i}ba, Brazil}
\email{cffarias,jroberto,petrov@fisica.ufpb.br}

\author{A.~C.~Lehum}
\affiliation{Escola de Ci\^encias e Tecnologia, Universidade Federal do Rio Grande do Norte\\
Caixa Postal 1524, 59072-970, Natal, Rio Grande do Norte, Brazil}
\email{andrelehum@ect.ufrn.br}

\author{J. R. Nascimento}
\affiliation{Departamento de F\'{\i}sica, Universidade Federal da Para\'{\i}ba\\
 Caixa Postal 5008, 58051-970, Jo\~ao Pessoa, Para\'{\i}ba, Brazil}
\email{cffarias,jroberto,petrov@fisica.ufpb.br}

\author{A. Yu. Petrov}
\affiliation{Departamento de F\'{\i}sica, Universidade Federal da Para\'{\i}ba\\
 Caixa Postal 5008, 58051-970, Jo\~ao Pessoa, Para\'{\i}ba, Brazil}
\email{cffarias,jroberto,petrov@fisica.ufpb.br}

\begin{abstract}
Within the superfield approach, we formulate the superfield supersymmetric aether-like Lorentz-breaking models, both in three- and in four-dimensional cases.
\end{abstract}

\maketitle

\section{Introduction}

The problem of the compatibility of the Lorentz symmetry breaking with the supersymmetry is one of the key problems in the development of the Lorentz-breaking field theories. Essentially, there are two ways to implement the supersymmetry into the Lorentz-breaking theories. The first way consists in the deformation of the structure of the supersymmetry generators and their algebra, and, consequently, of the supercovariant derivatives, this way has been originally proposed in \cite{KostBer}, with no extra superfields introduced. Some attempts to implement this approach on the tree level have been presented in the papers \cite{attempts} where the Lorentz symmetry breaking has been introduced through introducing of asymmetry between space and time derivatives, and some simple superfield calculations have been carried out.
The second way consists in introducing the additional superfields whose component expansion involves constant vectors (tensors) in the action while the structure of the supersymmetry  generators (and hence of their algebra, the corresponding supersymmetric covariant derivatives and of the superfields) is maintained to be the same (see f.e. \cite{Hel}), 

In this paper we essentially follow the first way, which allows to formulate the Lorentz-breaking deformation of the supersymmetry algebra in a systematic manner and does not require the introduction of the extra superfields.  Our aim, principally, consists not only in a detailed development of the superfield formalism for the Lorentz-breaking supersymmetric theories, but also in applying of powerful methodology of the superfield calculations of the effective potential (a great number of results obtained with use of this methodology  is presented in \cite{effpot,effpot1}) to these theories. The main attention will be given, first, to constructing the deformed supersymmetry, second, to calculating the effective potential for the simplest superfield models.

\section{Three-dimensional Lorentz-breaking deformation of the superspace}


In the usual three-dimensional spacetime the spinor representation relates the Lorentz group to $SL(2,\Re)$, therefore the fundamental representations acts on a Majorana two-component spinor, consequently the spinor supersymmetry generators $Q_\alpha$ are Hermitian. To extend the usual superpace to a three-dimensional deformed superspace, let us define the deformed supersymmetry generators as (cf. \cite{SGRS})
\begin{eqnarray}
Q_{\alpha}&=&i[\partial_{\alpha}-i\theta^{\beta}\gamma^m_{\beta\alpha}(\partial_m+k_{mn}\partial^n)]\nonumber\\
&=&i[\partial_{\alpha}-i{\theta}^{{\beta}}\gamma^m_{\beta\alpha}\nabla_m],
\end{eqnarray}
where $\partial_{\alpha}$ is the derivative with respect to the Grassmannian superspace coordinates $\theta^{\alpha}$, and $\nabla_m=\partial_m+k_{mn}\partial^n$ is a ``covariantized'' space-time derivative. The $k_{mn}$ is a constant tensor which implements the Lorentz symmetry breaking. Without restriction of generality, it can be chosen to assume an aether-like form $k_{mn}=\alpha u_mu_n$ \cite{aether}, where $u^m$ is a constant vector, and $\alpha$ is a some number, therefore we can refer to this algebra as to the aether-like generalization of the supersymmetry algebra, and denominate the theories constructed on its base as the  aether-like superfield supersymmetric Lorentz-breaking theories. However, we must prevent the reader that the methodology denominated in \cite{Sib}  as ``supersymmetric aether'', where the constant $u^m$ vector is used, represents itself as a supersymmetric extension of the Einstein-aether theory where the supersymmetry algebra is not deformed, but the $u^m$ is a lower component of the extra  dynamical (super)field, and thus has nothing common with our model. We note that the constant $k_{mn}$ is dimensionless, thus, its presence probably will not jeopardize the renormalizability of the field theory models.

The anticommutation relation between the deformed supersymmetry generators is
\bea
\{Q_{\alpha},Q_{\alpha}\}=2i\gamma^m_{{\alpha}\beta}\nabla_m,
\eea
that is an operator proportional to the simple space-time derivative, as it must be to satisfy the Leibnitz rule.

The new supercovariant derivative is constructed to anticommute with $Q_\alpha$, and it can be written as
\bea
\label{sder3d}
D_{\alpha}=\partial_{\alpha}+i{\theta}^{{\beta}}\gamma^m_{{\beta}\alpha}\nabla_m~,
\eea
where the operator $\nabla_m$ commutes with $D_{\alpha}$, as well as with the supersymmetry generators.

The supercovariant derivatives satisfy the relations:
\bea
&&\{D_{\alpha},{D}_{\beta}\}=2i\gamma^m_{\alpha\beta}\nabla_m~;\nonumber\\
&&(D^2)^2=\tilde{\Box}~;\nonumber\\
&&D^{\alpha}D_{\beta}D_{\alpha}=0,
\eea
where 
\bea
\tilde{\Box}=\nabla^m\nabla_m=\Box+2k_{mn}\partial^m\partial^n+k^{mn}k_{ml}\partial_n\partial^l
\eea
is a deformed d'Alembertian operator.

Just as in the usual case, we can define superfields over this deformed superspace and construct invariants over it. Let $\Phi$ be a {\emph{real}} scalar superfield. We can introduce a Wess-Zumino model whose action formally coincides with the usual one
\bea
\label{action3d}
S=\int d^5z\left[\frac{1}{2}{\Phi}(D^2+m)\Phi+\frac{\lambda}{6}\Phi^3\right],
\eea
while the structure of the superfields, however, is deformed. 

In general, superfields can be expanded in a Taylor series in the Grassmanian variable $\theta$ as
\begin{eqnarray}
\Phi(x,\theta)=\varphi(x)+\theta^\alpha\psi_\alpha(x)-\theta^2F(x)~.
\end{eqnarray}
But for our deformed superspace it is more convenient to define the superfield components by projection as 
\begin{eqnarray}
\varphi(x)&=&\Phi(x,\theta)\Big{|}_{\theta=0}~;\nonumber\\
\psi_\alpha(x)&=&D_\alpha\Phi(x,\theta)\Big{|}_{\theta=0}~;\\
F(x)&=&D^2\Phi(x,\theta)\Big{|}_{\theta=0}~,\nonumber
\end{eqnarray}
where the spinor supercovariant derivatives are given by Eq. (\ref{sder3d}). We note that since the ``deformed'' supercovariant derivatives differ from the usual ones only in the sector proportional to $\theta^{\alpha}$, the component contents of the superfields in ``deformed'' and usual cases will be exactly the same, thus, the ``deformed'' action in components looks like a sum of the usual action and some extra terms proportional to different degrees of the Lorentz-breaking parameters.
In this way, we can write the action (\ref{action3d}) in terms of the components as
\bea
\label{action3dc}
S&=&\int d^3x \Big[\frac{1}{2}F^2 +\frac{1}{2}\psi^\alpha i{(\gamma^m)_{\alpha}}^\beta \nabla_m\psi_\beta +\frac{1}{2}\varphi\tilde{\Box}\varphi +m(\psi^2+\varphi F)+\nonumber\\&+&
\lambda\left(\varphi\psi^2+\frac{1}{2}\varphi^2F\right)\Big].
\eea
So, we find that, for example, the free action for the fermion $S_f=\frac{1}{2}\int d^3x\psi^{\alpha}[i{(\gamma^m)_{\alpha}}^\beta \nabla_m+m]\psi_\beta=\frac{1}{2}\int d^3x\psi^{\alpha}[i{(\gamma^m)_{\alpha}}^\beta (\partial_m+k_{mn}\partial^n)+m]\psi_\beta$ acquires just the same additive aether-like term that was discussed in \cite{aether}. We note that the kinetic term for the scalar superfield $S_{sc,kin}=\frac{1}{2}\int d^3x\varphi\tilde{\Box}\varphi=\frac{1}{2}\int d^3x\varphi(\partial_m+k_{mn}\partial^n)(\partial^m+k^{ml}\partial_l)\varphi$, beside of the usual additive aether-like term $\varphi k^{mn}\partial_m\partial_n\varphi$ \cite{aether} involves also the extra higher-order term, which, for the case $k_{mn}=\alpha u_mu_n$, corresponds to the fourth degree of the $u_m$. 

Now we are able to obtain the Feynman rules for the three-dimensional aether superspace. Let us begin with the generating functional for the model defined by the action (\ref{action3d}), the Wess-Zumino model, with the adding of a source term. Let
\bea
\label{action3d1}
Z(J)&=&\int{\mathcal{D}}\Phi~\mathrm{exp}\Big{\{}\int d^5z\left[\frac{1}{2}{\Phi}(D^2+m)\Phi+\frac{\lambda}{6}\Phi^3+J\Phi\right]\Big{\}}\nonumber\\
&=&\mathrm{exp}\left[S_I\left(\frac{\delta}{\delta J}\right)\right]\int{\mathcal{D}}\Phi~\mathrm{exp}\Big{\{}\int d^5z\left[\frac{1}{2}{\Phi}(D^2+m)\Phi+J\Phi\right]\Big{\}}~,
\eea     
where $S_I(\Phi)=\frac{\lambda}{6}\int{d^5z}\Phi^3$.

Completing the square and performing the Gaussian integration over $\Phi$, we have 
\bea
\label{action3d2}
Z(J)=\mathrm{exp}\left[S_I\left(\frac{\delta}{\delta J}\right)\right]~\mathrm{exp}\Big{\{}-\frac{1}{2}\int d^5z~J\frac{1}{D^2+m}J\Big{\}}.
\eea   

Therefore, we can easily obtain the scalar superfield propagator in momentum space
\bea
\label{prop3d}
\langle \Phi(p,\theta_1)\Phi(-p,\theta_2)\rangle&=&\frac{(D^2-m)}{\tilde{p}^2+m^2}\delta^2(\theta_1-\theta_2)~,
\eea  
where $\tilde{p}^2=p^2+2k_{mn}p^mp^n+k^{mn}k_{ml}p_np^l$ and $D^2=\partial^2-\theta^\beta\gamma^m_{\beta\alpha}\tilde{p}_m\partial^\alpha +\theta^2\tilde{p}^2$.

We note that one can calculate the superficial degree of divergence of the corresponding Feynman supergraphs just in the same way as in the common superfield theories. Moreover, the result for it will coincide with the results obtained in the usual superfield theory since the propagators in undeformed and deformed theories have the same asymptotic behaviour, for example, for the scalar field theory the couplings $\Phi^3$ and $\Phi^4$ will again correspond to the renormalizable theories, and all theories except of those ones with exotic effective dynamics continue to be one-loop finite.

Let us discuss the dispersion relations in our theory. The denominator of (\ref{prop3d}) looks like $\tilde{p}^2+m^2=p^2+2k_{mn}p^mp^n+k^{mn}k_{ml}p_np^l+m^2$ (this structure is common for the propagators in the CPT-even Lorentz-breaking theories, see f.e. \cite{Fer}). Let us consider this denominator, for the signature $(-+++)$, and $k_{mn}=\alpha u_mu_n$, with $u_m=(u_0,\vec{u})$, and $u^mu_m\equiv \epsilon$ is equal either to 1, for the space-like $u_m$, or to $-1$, for the time-like $u_m$, or to 0, for the light-like $u_m$. 

1. Space-like $u_m$ case, $\epsilon=1$. We have $E^2=p^2+m^2+(2\alpha+\alpha^2)(\vec{u}\cdot\vec{p})^2$. We see that for both $\alpha>0$ and for $\alpha<0$, but $|\alpha|\ll 1$, the dynamics is consistent, where as for a negative $\alpha$ with a rather large absolute value the theory turns out to be degenerate or unstable. In particular, if $\alpha=-1$, and both the vector $\vec{u}$ and the vector $\vec{p}$ are directed along the $x$ axis, one has $E^2=m^2$, so, the dynamics is degenerate. We note that it is just the case when the matrix $S_{mn}=\eta_{mn}+k_{mn}$ is degenerate, thus, the degeneracy of the matrix $S_{mn}$ results in the degeneracy of the dynamics.

2. Time-like $u_m$ case, $\epsilon=-1$. We have $E^2(1-\alpha)^2=\vec{p}^2+m^2$. So, the dynamics is consistent everywhere except of the case $\alpha=1$ which signalizes the degeneracy of the matrix $S_{mn}$.

3. Light-like case, $\epsilon=0$, and $u_m=(1,1,0,0)$. In this case, when both the vector $\vec{u}$ (the spatial part of $u_m$) and the vector $\vec{p}$ are directed along the $x$ axis, we have $E=\frac{1}{1-2\alpha}[-2\alpha p\pm\sqrt{p^2(1+2\alpha+4\alpha^2)+m^2}]$, so, if $|\alpha|\ll 1$, the dynamics is consistent.

In principle, one can also point out the case when the dispersion relations are not modified even for the nontrivial $u_m$, that is, the case $\alpha u^mu_m+2=0$, which yields $\tilde{p}^2=p^2$, but this situation is impossible if we have $|\alpha|\ll 1$, while $u^mu_m$ is restricted to have values $-1,0,1$ only. 

Thus, we conclude that if we impose the condition $|\alpha|\ll 1$, together with $u^mu_m$ is either $\pm 1$ or 0, to ensure that the Lorentz-breaking terms can be treated as a small correction, hence the quadratic form corresponding to $\tilde{p}^2$ is never degenerate. 

The analysis in the four-dimensional theories which will be considered later in this paper, is just the same, because of the same structure of the denominator of the propagator. In principle, it is natural to expect that the similar situation will take place in all theories where the Lorentz symmetry breaking is introduced through a deformation of the supersymmetry algebra as in this paper.

Now, let us compute the quadratic part of the effective action for the three-dimensional Wess-Zumino model at one-loop level. The Feynman diagram that contribute with the process is depicted in Fig. 1. 

\begin{figure}[ht]
\centerline{\includegraphics{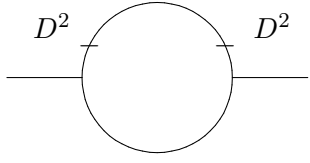}} 
\caption{The contribution to the kinetic term in the three-dimensional superspace.}
\end{figure}

The corresponding expression can be cast as 
\begin{eqnarray}
\Gamma^{(1)}&=&\frac{\lambda^2}{6}\int{\frac{d^3p}{(2\pi)^3}d^2\theta}~\Phi(-p,\theta)\left(D^2-2m\right)\Phi(p,\theta)
\times\nonumber\\&\times&
\int{\frac{d^3q}{(2\pi)^3}}\frac{1}{(\tilde{q}^2+m^2)[(\tilde{q}+\tilde{p})^2+m^2]}~.
\end{eqnarray}

It is clear that this expression is finite, and, if the external momentum vanishes, $p=0$, one can
change the variables and the  integration measure as $\int d^3q=\Delta\int d^3\tilde{q}$, $\Delta=\det(\frac{\partial q^m}{\partial\tilde{q}^n})=\det^{-1}(\delta^m_n+k^m_n)$ is a Jacobian of change of variables, it is a constant (in the case of the small $u^a$, one has $\Delta=1-u^2$). So, one arrives at
\bea
\Gamma^{(1)}=\Delta\frac{\lambda^2}{6}\int{\frac{d^3p}{(2\pi)^3}d^2\theta}~\Phi(-p,\theta)\left(D^2-2m\right)\Phi(p,\theta)
\frac{1}{8\pi|m|}.
\eea
We conclude that this methodology does not essentially differ from the usual supergraph technique. It can be naturally generalized for more sophisticated theories, in particular, the gauge ones.

\section{Four-dimensional Lorentz-violating deformation of the superspace}

We start with the following deformation of the supersymmetry generators \cite{KostBer} in the four-dimensional case (cf. \cite{BK0,SGRS}):
\bea
Q_{\alpha}=\partial_{\alpha}-i\bar{\theta}^{\dot{\beta}}\sigma^m_{\dot{\beta}\alpha}(\partial_m+k_{mn}\partial^n);\nonumber\\
\bar{Q}_{\dot{\alpha}}=\partial_{\dot{\alpha}}-i\theta^{\beta}\bar{\sigma}^m_{\beta\dot{\alpha}}(\partial_m+k_{mn}\partial^n).
\eea
Here $\partial_{\alpha}$, $\partial_{\dot{\alpha}}$ are the simple derivatives with respect to the Grassmannian superspace coordinates $\theta^{\alpha}$, $\bar{\theta}^{\dot{\alpha}}$, that is, $\partial_{\alpha}=\frac{\partial}{\partial\theta_{\alpha}}$, and  $\partial_{\dot{\alpha}}=\frac{\partial}{\partial\bar{\theta}_{\dot{\alpha}}}$. These generators are linear in the derivatives, as it must be to satisfy the Leibnitz rule. Just as in the three-dimensional case, the $k_{mn}$ is a constant tensor implementing the Lorentz symmetry breaking. Again, without restrictions on the generality, we can choose it to be symmetric, f.e. $k_{mn}=\alpha u_mu_n$, with $u^m$ is a constant vector, and $|\alpha|\ll 1$, that is, in the aether-like form. 

Effectively, these generators can be represented as
\bea
Q_{\alpha}=\partial_{\alpha}-i\bar{\theta}^{\dot{\beta}}\sigma^m_{\dot{\beta}\alpha}\nabla_m;\nonumber\\
\bar{Q}_{\dot{\alpha}}=\partial_{\dot{\alpha}}-i\theta^{\beta}\bar{\sigma}^m_{\beta\dot{\alpha}}\nabla_m,
\eea
The anticommutation relation of the supersymmetry generators is
\bea
\{Q_{\alpha},\bar{Q}_{\dot{\alpha}}\}=-2i\sigma^m_{\dot{\alpha}\alpha}\nabla_m,
\eea
so, it indeed gives an operator proportional to the translation as it must be in the supersymmetric field theories (cf. \cite{BK0,SGRS}).

The corresponding supercovariant derivative must anticommute with these generators, being
\bea
\label{sder}
D_{\alpha}=\partial_{\alpha}+i\bar{\theta}^{\dot{\beta}}\sigma^m_{\dot{\beta}\alpha}\nabla_m;\nonumber\\
\bar{D}_{\dot{\alpha}}=\partial_{\dot{\alpha}}+i\theta^{\beta}\bar{\sigma}^m_{\beta\dot{\alpha}}\nabla_m.
\eea
It is clear that the operator $\nabla_m$ commutes with $D_{\alpha}$, $\bar{D}_{\dot{\alpha}}$, as well as with the supersymmetry generators.

By analogy with the usual four-dimensional superfield supersymmetry \cite{BK0} (throughout this section, we use the normalization relations for the supersymmetry generators and supercovariant derivatives which in the Lorentz-invariant case are reduced to those ones from \cite{BK0}) one can show that supercovariant derivatives satisfy the relations: 
\bea
\{D_{\alpha},\bar{D}_{\dot{\alpha}}\}=2i\sigma^m_{\dot{\alpha}\alpha}D_m;\nonumber\\
D^2\bar{D}^2D^2=16\tilde{\Box}D^2;\nonumber\\
D_{\alpha}D_{\beta}D_{\gamma}=\bar{D}_{\alpha}\bar{D}_{\beta}\bar{D}_{\gamma}=0.
\eea
We introduce the chiral superfield $\Phi$ satisfying the relation $\bar{D}_{\dot{\alpha}}\Phi=0$, and the antichiral one $\bar{\Phi}$ satisfying the relation $D_{\alpha}\bar{\Phi}=0$, where the supercovariant derivatives are "new" ones satisfying the relations (\ref{sder}). 

As a simplest example, we can introduce the Lorentz-breaking Wess-Zumino model whose action formally coincides with the usual one
\bea
S=\int d^8z \Phi\bar{\Phi}+[\int d^6z(\frac{m}{2}\Phi^2+\frac{\lambda}{3!}\Phi^3)+h.c.],
\eea
while the structure of the superfields, however, is deformed.

Let us introduce the component structure of the chiral superfield via projections (cf. \cite{SGRS}; however, here we use the normalizations for supercovariant derivatives adopted in \cite{BK0}):
\bea
&&\phi=\Phi|_{\theta=\bar{\theta}=0};\nonumber\\
&&\psi_{\alpha}=\frac{1}{2}D_{\alpha}\Phi|_{\theta=\bar{\theta}=0};\nonumber\\
&&F=\frac{D^2}{4}\Phi|_{\theta=\bar{\theta}=0},
\eea
with the analogous definitions for the components of the antichiral field.
Thus, the component expansion of the Wess-Zumino model takes the form
\bea
S=\int d^4x \Big[\phi \tilde{\Box}\bar{\phi}+\psi^{\alpha}i\sigma^m_{\alpha\dot{\alpha}}\nabla_m\bar{\psi}^{\dot{\alpha}}+F\bar{F}+\lambda(\phi\psi^{\alpha}\psi_{\alpha}+\frac{1}{2}\phi^2 F+h.c.)
\Big].
\eea
We see that again the aether terms arise for the scalar and spinor component fields, and a fourth-order term arises for the scalar field. 

The propagators of the superfields have the same form as in the usual Wess-Zumino model, but with the modified d'Alembertian operator since the structure of covariant derivatives is modified:
\bea
<\Phi(z_1)\bar{\Phi}(z_2)>&=&\frac{1}{\tilde{\Box}-m^2}\delta(z_1-z_2);\\
<\Phi(z_1)\Phi(z_2)>&=&\frac{mD^2}{4\tilde{\Box}(\tilde{\Box}-m^2)}\delta(z_1-z_2), \quad\,
<\bar{\Phi}(z_1)\bar{\Phi}(z_2)>=\frac{m\bar{D}^2}{4\tilde{\Box}(\tilde{\Box}-m^2)}\delta(z_1-z_2),\nonumber
\eea
with the additional D-factors are associated with the vertices by the same rule as in the usual supersymmetric models. Since each modified d'Alembertian operator $\tilde{\Box}$ is of the second order in the space-time derivatives just as the usual d'Alembertian operator, the calculation of the superficial degree of divergence $\omega$ does not differ  from the usual case yielding
\bea
\omega=2-E-C,
\eea
where $E$ is a number of external legs, and $C$ is a number of $<\Phi\Phi>$, $<\bar{\Phi}\bar{\Phi}>$ propagators.
We see that there is only one type of the divergences in this theory, that is, the divergent correction to the kinetic $\Phi\bar{\Phi}$ term depicted by the graph depicted at Fig. 2.

\begin{figure}[ht]
\centerline{\includegraphics{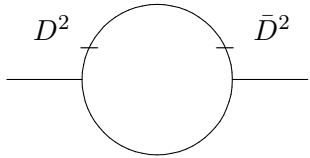}} 
\caption{The contribution to the kinetic term in the four-dimensional superspace.}
\end{figure}

\noindent After the simple D-algebra transformations (we note that shrinking any loop into a point is based on the same identity $\delta_{12}\bar{D}^2D^2\delta_{12}=16\delta_{12}$ as in the usual case), we have the kinetic term in the form
\bea
\Gamma_2=\frac{\lambda^2}{2}\int d^4\theta\int\frac{d^4p}{(2\pi)^4}\bar{\Phi}(-p,\theta)\Phi(p,\theta)\int\frac{d^4q}{(2\pi)^4}
\frac{1}{(\tilde{q}^2-m^2)((\tilde{q}+\tilde{p})^2-m^2)}.
\eea
This contribution is evidently logarithmically divergent. To evaluate it, we carry out the Wick rotation and Feynman parameterization which yield
\bea
\Gamma_2=i\frac{\lambda^2}{2}\int d^4\theta\int\frac{d^4p}{(2\pi)^4}\bar{\Phi}(-p,\theta)\Phi(p,\theta)\int_0^1 dx\int\frac{d^4q}{(2\pi)^4}
\frac{1}{(\tilde{q}^2+m^2-\tilde{p}^2x(1-x))^2}.
\eea
Then, we change the variables and the  integration measure in the same way as above, i.e. $\int d^4q=\Delta\int d^4\tilde{q}$, $\Delta=\det(\frac{\partial q^m}{\partial\tilde{q}^n})=\det^{-1}(\delta^m_n+k^m_n)$ is a Jacobian of change of variables, it is a constant (in the case of the small $u^a$, one has $\Delta=1-u^2$). As a result, we get
\bea
\Gamma_2=i\Delta\frac{\lambda^2}{2}\int d^4\theta\int\frac{d^4p}{(2\pi)^4}\bar{\Phi}(-p,\theta)\Phi(p,\theta)\int_0^1 dx\int\frac{d^4\tilde{q}}{(2\pi)^4}
\frac{1}{(\tilde{q}^2+m^2+\tilde{p}^2x(1-x))^2}.
\eea
The integral is the same as in the usual Wess-Zumino model case \cite{BK0,SGRS}, and we have
\bea
\Gamma_2=i\Delta\frac{\lambda^2}{2}\int d^4\theta\int\frac{d^4p}{(2\pi)^4}\bar{\Phi}(-p,\theta)\Phi(p,\theta)(\frac{1}{16\pi^2\epsilon}+\int_0^1 dx\ln\frac{m^2+\tilde{p}^2x(1-x)}{\mu^2}),
\eea
so, the finite part is not Lorentz-invariant being dependent on the Lorentz-noninvariant object $\tilde{p}^2$.

Then, we can calculate the one-loop K\"{a}hlerian effective potential \cite{BK0}, which, after carrying out the same steps as in the usual case (see \cite{effpot1} for the details), yields
\bea
K^{(1)}=-\frac{1}{2}\sum\limits_{n=1}^{\infty}\int d^8z[\Psi\bar{\Psi}\frac{D^2\bar{D^2}}{16\tilde{\Box}^2}]\delta^8(z-z')|_{z=z'},
\eea
where $\Psi=m+\lambda\Phi$, $\bar{\Psi}=m+\lambda\bar{\Phi}$. This expression yields
\bea
K^{(1)}=-\frac{1}{32\pi^2}\int d^4\theta\int\frac{d^4q}{(2\pi)^4}\bar{\Psi}\Psi\ln(1-\frac{\bar{\Psi}\Psi}{(q_m+k_{mn}q^n)^2}).
\eea
This expression can be integrated. To do it, we again carry out the change of variables $q_m+k_{mn}q^n\to \tilde{q}_m$, and arrive at
\bea
K^{(1)}=-\frac{1}{32\pi^2}\Delta\int d^4\theta\int\frac{d^4\tilde{q}}{(2\pi)^4}\bar{\Psi}\Psi\ln(1-\frac{\bar{\Psi}\Psi}{\tilde{q}^2}),
\eea
where $\Delta$ is the same as above. After Wick rotation and integration proceeded in the same way as in \cite{effpot1}, together with the subtracting of the corresponding counterterm (which differs from the usual counterterm used in the Wess-Zumino model \cite{BK0} only by a constant factor $\Delta$) we arrive at
\bea
K^{(1)}=-\frac{1}{32\pi^2}\Delta\Psi\bar{\Psi}\ln\frac{\Psi\bar{\Psi}}{\mu^2}.
\eea
Thus, we see that the contribution to the k\"{a}hlerian effective potential does not essentially differ from that one in the Wess-Zumino model. 

\section{Summary}

Basing on the Berger-Kostelecky construction \cite{KostBer}, we developed the superfield approach for constructing the Lorentz-breaking supersymmetric field theories. This approach turns out to be no more complicated as the standard supergraph technique whose examples of application are present in the papers \cite{effpot}, and the results do not crucially differ from the usual case. We note that if the deformation of the supersymmetry algebra is small, the dynamics continues to be consistent. It is interesting to note that, first, this scheme is essentially CPT-even, second, in principle, one can choose the Lorentz-breaking matrix $k_{mn}$ to be antisymmetric, and in this case the integral measure is not corrected for the infinitesimal $k_{mn}$, $\Delta=1$. In other words, we succeeded to conciliate Lorentz symmetry breaking and the supersymmetry in a rather simple way. In principle, the calculations for the supergauge theories can be carried out in the same way. We are planning to do it in a forthcoming paper.

{\bf Acknowledgments.}
This work was partially supported by Conselho Nacional de Desenvolvimento Cient\'{\i}fico e Tecnol\'{o}gico (CNPq) and Funda\c{c}\~{a}o de Apoio \`{a} Pesquisa do Estado do Rio Grande do Norte (FAPERN). The work by A. C. Lehum and A. Yu. P. has been supported by the CNPq project No. 303392/2010-0 and 303461/2009-8, respectively.

\end{document}